\documentclass[sigconf]{acmart}

\usepackage[nolist]{acronym}
\AtBeginDocument{%
 \providecommand\BibTeX{{%
    \normalfont B\kern-0.5em{\scshape i\kern-0.25em b}\kern-0.8em\TeX}}}

\usepackage{soul}
\usepackage{color}
\usepackage{wallpaper}
\usepackage{graphicx}
\usepackage{cleveref}
\usepackage{numprint}
\usepackage{paralist}
\usepackage{xspace}
\usepackage[lofdepth,lotdepth]{subfig}

\newcommand{\cellbg}[1]{\cellcolor{lightgray}\textbf{#1}}
\usepackage{colortbl}

\definecolor{orangeX}{rgb}{1,.5,0}
\definecolor{blueX}{rgb}{.2, .59, .88}
\definecolor{purpleX}{rgb}{.294118, 0, .509804}
\definecolor{greenX}{rgb}{.421, .578, .241}
\definecolor{bole}{rgb}{0.47, 0.27, 0.23}
\definecolor{mypink3}{cmyk}{0, 0.7808, 0.4429, 0.1412}
\definecolor{mygray}{gray}{0.6}
	
\newcommand{\audeep}{\mbox{\textsc{auDeep}}}

\usepackage{arydshln}
\makeatletter
\def\adl@drawiv#1#2#3{%
        \hskip.5\tabcolsep
        \xleaders#3{#2.5\@tempdimb #1{1}#2.5\@tempdimb}%
                #2\z@ plus1fil minus1fil\relax
        \hskip.5\tabcolsep}
\newcommand{\cdashlinelr}[1]{%
  \noalign{\vskip\aboverulesep
           \global\let\@dashdrawstore\adl@draw
           \global\let\adl@draw\adl@drawiv}
  \cdashline{#1}
  \noalign{\global\let\adl@draw\@dashdrawstore
           \vskip\belowrulesep}}
\makeatother

\acrodef{CNN}[CNN]{Convolutional Neural Network}
\acrodef{GRU}[GRU]{Gated Recurrent Unit}
\acrodef{UAR}[UAR]{Unweighted Average Recall}

\newcommand{\ds}{\textsc{DeepSpectrum}}

\newcommand{\eg}{e.\,g., }

\copyrightyear{2023}
\acmYear{2023}
\setcopyright{acmlicensed}\acmConference[MM '23]{Proceedings of the 31st ACM International Conference on Multimedia}{October 29--November 3, 2023}{Ottawa, Canada}
\acmBooktitle{Proceedings of the 31st ACM International Conference on Multimedia (MM '23), October 29--November 3, 2023, Ottawa, Canada}
\acmPrice{15.00}
\acmDOI{10.1145/3503161.3551591}
\acmISBN{978-1-4503-9203-7/22/10}

\begin{document}
\title[ComParE 2023 Baseline Paper]{
The ACM Multimedia 2023 Computational \\Paralinguistics Challenge: Emotion Share \& Requests} 

\author{Bj\"orn W.\ Schuller}
\affiliation{%
  \institution{GLAM, Imperial College London}
  \city{London}
  \country{United Kingdom}}

\author{Anton Batliner}
\affiliation{%
  \institution{EIHW, University of Augsburg}
  \city{Augsburg}
  \country{Germany}}

\author{Shahin Amiriparian}
\affiliation{%
  \institution{EIHW, University of Augsburg}
  \city{Augsburg}
  \country{Germany}}

\author{Alexander Barnhill}
\affiliation{%
  \institution{FAU}
  \city{Erlangen-Nuremberg}
  \country{Germany}}

\author{Maurice Gerczuk}
\affiliation{%
  \institution{EIHW, University of Augsburg}
  \city{Augsburg}
  \country{Germany}}

\author{Andreas Triantafyllopoulos}
\affiliation{%
  \institution{EIHW, University of Augsburg}
  \city{Augsburg}
  \country{Germany}}

\author{Alice Baird}
\affiliation{%
  \institution{Hume AI}
  \city{New York}
  \country{USA}}

\author{Panagiotis Tzirakis}
\affiliation{%
  \institution{Hume AI}
  \city{New York}
  \country{USA}}

\author{Chris Gagne}
\affiliation{%
  \institution{Hume AI}
  \city{New York}
  \country{USA}}

\author{Alan S.\ Cowen}
\affiliation{%
  \institution{Hume AI}
  \city{New York}
  \country{USA}}

\author{Nikola Lackovic}
\affiliation{%
  \institution{Malakoff Humanis}
  \city{Paris}
  \country{France}}

\author{Marie-José Caraty}
\affiliation{%
  \institution{Sorbonne University}
  \city{Paris}
  \country{France}}

\author{Claude Montaci\'e}
\affiliation{%
  \institution{Sorbonne University}
  \city{Paris}
  \country{France}}

\renewcommand{\shortauthors}{Bj\"orn W.\ Schuller et al.}
\begin{abstract}
The ACM Multimedia 2023 Computational Paralinguistics Challenge addresses two different problems for the first time in a research competition under well-defined conditions: In the \textit{Emotion Share} Sub-Challenge, a regression on speech has to be made; and in the \textit{Requests} Sub-Challenges, \textit{requests} and \textit{complaints} need to be detected. We describe the Sub-Challenges, baseline feature extraction, and classifiers based on the `usual' \textsc{ComParE} features, the \audeep{} toolkit, and deep feature extraction from pre-trained CNNs using the \ds{} toolkit; in addition, 
 wav2vec2 models are used.
\end{abstract}



\maketitle
\section{Introduction}

	In this ACM Multimedia 2023 \textsc{COMputational PARalinguistics  challengE  (\textsc{ComParE})} -- the 15th since 2009 \cite{Schuller09-TI2,Schuller11-RRE},
	we address two new problems within the field of Computational Paralinguistics \cite{Schuller14-CPE} in a challenge setting:

\subsection{Emotion Share Sub-Challenge}

As the ComParE challenge has been making efforts to address for the past decade~\cite{Schuller09-TI2}, the lack of data in the domain of emotion modelling continues to be an issue for real-world development, particularly with the rise of data-hungry deep learning methods~\cite{batliner2020ethics}.  
To address this, for the \textbf{Emotion Share Sub-Challenge}, Hume AI has provided the Hume-Prosody dataset. The Sub-Challenge features a multi-label regression task. For each of the nine different emotions, a proportion or `share'  has been assigned to the nine emotions based on the proportion of raters that rated that emotion for the `seed' sample. For further details on the data collection methodology,  see \cite{Cowen19-M2E} and Section \ref{Emotion-Share-description}.

\subsection{Requests Sub-Challenge}

The interaction between an organisation and its customers (Customer  Relationship  Management CRM) \cite{Anshari19-CRM} takes often place within  a call centre \cite{Whiting06-MVE}, \eg via a phone call. Speech analysis \cite{Scheidt19-MAC,Hildebrand20-VAI} helps to model this interaction and the interest of the users as well as to pinpoint problems. 
In the \textbf{Requests Sub-Challenge}, we are interested in two 
tasks  related  to  CRM:  the  classification  of  customer  \emph{Requests} 
 and  
 \emph{Complaints}.   
 Data and annotations have been  provided by 
 STIH Laboratory, Sorbonne University.

\subsection{Tasks}

For these two challenges, 
either a target class has to be predicted (for Request and Complaints), or a correlation measure has to be computed (for Emotion Share).
Contributors can employ their own features and machine learning (ML) algorithms; standard feature sets and procedures are provided.
Participants have to use the pre-defined
partitions for each Sub-Challenge. 
They may report results that they  
obtain from the \textbf{Train}(ing)/\textbf{Dev}(elopment) 
set 
but have only 
five trials to upload their results on the \textbf{Test} set per Sub-Challenge, whose labels are unknown to them. 	
Each participation must be accompanied by a paper presenting the results, which undergoes peer-review.
The organisers preserve the right to re-evaluate the findings, but will not participate in the Challenge. 
	As evaluation measure, 
    we employ  the \textbf{Unweighted Average Recall (UAR)} 
    for the   
    Requests Sub-Challenges as used since the first Challenge from 2009 \cite{Schuller09-TI2,Schuller11-RRE};
    it is more adequate for (unbalanced) multi-class classifications than Weighted Average Recall (i.\,e., accuracy)
   \cite{Schuller14-CPE,Rosenberg12-CSD}.
      For the Emotion-Share Sub-Challenge, Spearman's $\rho$  \cite{Spearman04-TPA} is used as most adequate measure for such ranking values.
	Ethical approval for the studies has been obtained. %

	\section{The Two Sub-Challenges}
	\label{Corpora}

\subsection{Emotion Share -- The \textbf{Hume-Prosody} Corpus \textbf{HP-C}}
\label{Emotion-Share-description}    

The basis for the dataset is more than 5,000 `seed' samples. Seeds consist of various emotional expressions  (e.\,g., 
\textit{`Tal vez, sea verdad'}), which were gathered from openly available datasets including  MELD \cite{poria2018meld} and  VENEC \cite{Laukka10-PTV,elfenbein2022we,laukka2016expression}.  The seeds are a mixture of `same' and `different' sentences -- \eg  more than 500 instances of \textit{`Let me tell you something'} \cite{Laukka10-PTV}, where the functional load of prosody is high, and `different' sentences (each of them with different words and semantics), where the functional load of prosody is lower.

The seed samples were mimicked by speakers recruited via Amazon Mechanical Turk  \cite{cowen2019primacy}. The Sub-Challenge subset consists of 51,881 `mimic' samples (total of 
41:48:55\,h of data,   
mean 2.9\,s., range 1.2 -- 7.98\,s) from 1,004 speakers aged from 20 to 66\,years old. It was gathered in 3 countries with broadly differing cultures: the United States, South Africa, and Venezuela.
For data processing, files below 1\,s and above 8\,s were excluded. 
The data were recorded at home via the speakers' microphones. The full Hume-Prosody dataset consists of 48 dimensions of emotional expression and is based on the semantic-space model for emotion~\cite{cowen2021semantic}. For this Sub-Challenge, nine emotional classes  have been selected due to their more balanced distribution across the valence-arousal space: `Anger', `Boredom', `Calmness', `Concentration', `Determination', `Excitement', `Interest', `Sadness', and `Tiredness'. 

Each seed sample was rated by the individual that imitated it using a `select-all-that-apply' method \cite{cowen2019primacy}. 
Seeds were assigned a mean of 2.03\,emotions per rater (max: 7.11, min: 1.00), with a standard deviation of 1.33\,emotions.
The proportion of times a given seed sample was rated with each emotion was then applied to all mimics of that seed sample. This results in the \textbf{share} per emotion assigned by the speakers.
For the Sub-Challenge baseline, the labels have been scaled to a maximum of 1 by dividing by the maximum emotion value per sample across the nine emotions. 

\subsection{Requests -- The HealthCall30 Corpus \textbf{HC-C}: Complaints and Requests}

This is a subset (audio-only) of the HealthCall30 corpus, provided by Montacié and colleagues \cite{Lackovic22-POU}. It is based on real audio interactions between call centre agents and customers who call to solve a problem or to request information. This corpus is designed to study natural spoken conversations and to predict Customer Relationship Management (CRM) annotations made by human agents from various vocal interaction, audio, and linguistic features. The corpus consists of 13,409 chunks of spoken conversations, each lasting 30 seconds. Two different classifications of these conversations were performed by annotators, based on CRM annotations:  the presence of customer complaints (\emph{Complaint} -- \textit{``yes''} or \textit{``no''}) and the type of customer request (\emph{Request}) -- either concerning membership issues (\textit{affil}) or a process (\textit{presta}), e.\,g., a reimbursement.
Each conversation was recorded on two separate and distinct audio channels; the first channel corresponds to the customer’s audio, and the second corresponds to the agent’s audio. More information can be found in \cite{Lackovic22-POU}. For the challenge, we provide both the raw dual channel audio files, as well as the normalised mono-conversions, as utilised for feature extraction and wav2vec2 training in the baseline.

\begin{table*}[t!]
\caption{Summary of the databases presented per Sub-Challenge. Number of instances per class in the Train/Dev/Test splits. 
Test split distributions for HC-C Requests and HC-C Complaints  are blinded during the ongoing challenge.
}
\label{tab:db}
\centering
\begin{tabular}{lrrrrlrrrrlrrrr} \toprule
\multicolumn{5}{c}{\textbf{HC-C Requests}: classification task (\#)}  & \multicolumn{5}{c}{\textbf{HC-C Complaints}: classification task (\#)}  & \multicolumn{5}{c}{\textbf{HP-C}: regression task} \\ 
\cmidrule(lr){1-5}\cmidrule(lr){6-10}\cmidrule(lr){11-15}
Class       & Train & Dev   & Test  & $\Sigma$  &   Class   & Train & Dev   & Test  & $\Sigma$  &           & Train & Dev   & Test  & $\Sigma$  \\ 
\midrule
affil       & 3,690  & 1,552  & -----  & -----     &   Yes     & 2,522  & 1,131  & -----  & -----       &   sample no.    & 30,133  & 12,241  & 9,507  & 51,881     \\
presta      & 3,132  & 1,532  & -----  & -----     &   No      & 4,300  & 1,953  & -----  & -----       &   speaker no.   & 600     &    202  &   202  & 1,004     \\
\cdashlinelr{1-10}
$\Sigma$    & 6,822  & 3,084  & 3,503  & 13,409    & $\Sigma$  & 6,822  & 3,084  & 3,503  & 13,409      & gender (f:m)    & 379:221 & 117:85  & 141:61  & 637:367     \\
\bottomrule 
\end{tabular}
\end{table*}

\section{Experiments and Results}
\label{experiments}
\vspace{-0.1cm}

%
\Cref{tab:db} shows the number of 
data for Train, Dev, and Test for the different 
corpora.

\begin{table*}[t!]
    \caption{Results for the Sub-Challenges. The official best results for Test yielding the \textbf{official baselines}  are highlighted (bold and greyscale); there are \textbf{no} official baselines for Dev.  \textbf{UAR}: Unweighted Average Recall. CI on Test: Confidence Intervals on Test, see explanation in the text. } 
    \centering
    \begin{tabular}{lccccccccc}
    \toprule
    \relax [UAR \%] & \multicolumn{3}{c}{\textbf{HC-C Request}} & \multicolumn{3}{c}{\textbf{HC-C Complaint}} & \multicolumn{3}{c}{\textbf{HP-C} [$\rho$]} \\
    \cmidrule(lr){2-4} \cmidrule(lr){5-7} \cmidrule(lr){8-10}
   Approach         & Dev   & Test  & CI on Test    & Dev   & Test  & CI on Test    & Dev   & Test  & CI on Test    \\
   \midrule
    
    wav2vec2        & 65.1  & \cellbg{67.2}  & 65.7 -- 68.7  & 50.9  & 52.2  & 50.7 -- 53.7  & .500  & .\cellbg{514}  & .499 -- .529  \\

    auDeep          & 54.9  & 52.4  & 50.7 -- 54.0  & 50.9  & 50.9  & 49.0 -- 52.7  & .347  & .357  & .341 -- .374  \\

    DeepSpectrum    & 58.2  & 55.6  & 54.0 -- 57.2  & 49.6  & 51.7  & 50.1 -- 53.4  & .335  & .331  & .313 -- .349  \\
    
    ComParE         & 60.9  & 58.8  & 57.2 -- 60.5  & 51.8  & \cellbg{52.9}  & 51.2 -- 54.7  & .359  & .365  & .347 -- .382  \\
    \cdashlinelr{1-10}
    Late Fusion     & 60.5  & 59.5  & 57.8 -- 61.2  & 51.2  & 51.8  & 50.1 -- 53.5  & .470  & .476  & .461 -- .492  \\
    \bottomrule
    \end{tabular}
    \label{tab:results}
\end{table*}

	\subsection{Approaches}
	\label{ssec:approaches}
This year, we evaluate four baseline systems. Of those, three consist of a distinct feature extraction step followed by a linear \ac{SVM} while the last system 
employes a pre-trained wav2vec2 model, taking raw audio as input. For a comprehensive view of the chosen hyperparameters, the reader is referred to the official baseline repository.

\subsubsection{\textbf{\textsc{ComParE} Acoustic Feature Set: }}
The official baseline feature set from openSMILE is the same as has been used in previous editions of the \textsc{ComParE} challenges, starting from 2013~\cite{Schuller13-TI2,Eyben13-RDI}.

 \subsubsection{\textbf{\ds{}:}}   
    
 This toolkit\footnote{\url{https://github.com/DeepSpectrum/DeepSpectrum}}   is applied to obtain deep representations from the input audio data utilising image pre-trained \acp{CNN}~\cite{Amiriparian17-SSC,amiriparian2019deep}. 
 has been used in previous challenges \cite{schuller2020interspeech,schuller2021interspeech} and is described in  \cite{Amiriparian17-SSC}. The efficacy of \ds{} features have been demonstrated for speech and audio recognition tasks~\cite{Amiriparian20-TCP}. For this iteration of the challenge, we utilise DenseNet169 to extract features from Mel-spectrograms with 128 bands.

 \subsubsection{\textbf{\audeep{}:}}   

This toolkit\footnote{\url{https://github.com/auDeep/auDeep}}     is obtained through unsupervised representation learning with recurrent sequence-to-sequence autoencoders
\cite{Amiriparian17-STS,Freitag18-AUL}. We choose a two-layer architecture utilising \ac{GRU} cells with $256$ hidden units and a latent feature size of $1024$ and train it for 64 epochs on Mel-spectrograms with 128 Mel-bands. Four variants of this autoencoder are trained, each clipping out low amplitudes from the input signals below -30\,dB, -45\,dB, -60\,dB and -75\,dB, respectively. The final features are obtained by concatenating the hidden representations of these four autoencoders. 


 \subsubsection{\textbf{\ac{SVM}:}}   
\label{sec:svm}
We train and evaluate \acp{SVM} with linear kernels on the three feature sets described above. The choice of feature normalisation, either min-max scaling or normalisation to zero mean and unit variance, is jointly optimised with the cost parameter $C$ of the \ac{SVM} based on the performance on the Dev set. After this optimisation, a final model is fit on the concatenated training and Dev sets for evaluation on the Test partition. 

 \subsubsection{\textbf{{Wav2Vec2}:}}   
 
For HC-C Requests and HC-C Complaints, we fine-tune a pre-trained Wav2Vec2 model for the challenge tasks on the raw audio files. We obtain the weights for the XLSR pre-trained model from huggingface hub~\footnote{\href{https://huggingface.co/jonatasgrosman/wav2vec2-large-xlsr-53-english}{https://huggingface.co/jonatasgrosman/wav2vec2-large-xlsr-53-english}}. We freeze the convolutional feature extractor and add final layers for classification to the output of the Transformer encoder. The model is trained for a maximum of $15$ epochs with a batch size of $64$ and learning rate of $3e^{-4}$. The best model, measured by \ac{UAR} on the development set, is saved and restored for the final evaluation on the test partition. For Emotion share, no fine-tuning for the challenge tasks was done -- instead, we use a model fine-tuned on the MSP-Podcast~\cite{Lotfian_2019_3} dataset\footnote{\href{https://huggingface.co/audeering/wav2vec2-large-robust-12-ft-emotion-msp-dim}{https://huggingface.co/audeering/wav2vec2-large-robust-12-ft-emotion-msp-dim}} to extract features; these are then fed to an SVM following the same procedure as detailed in \Cref{sec:svm}.

\begin{figure}[t!]
    \subfloat[HC-C Requests Confusion Matrix]{\includegraphics[width=0.49\columnwidth]{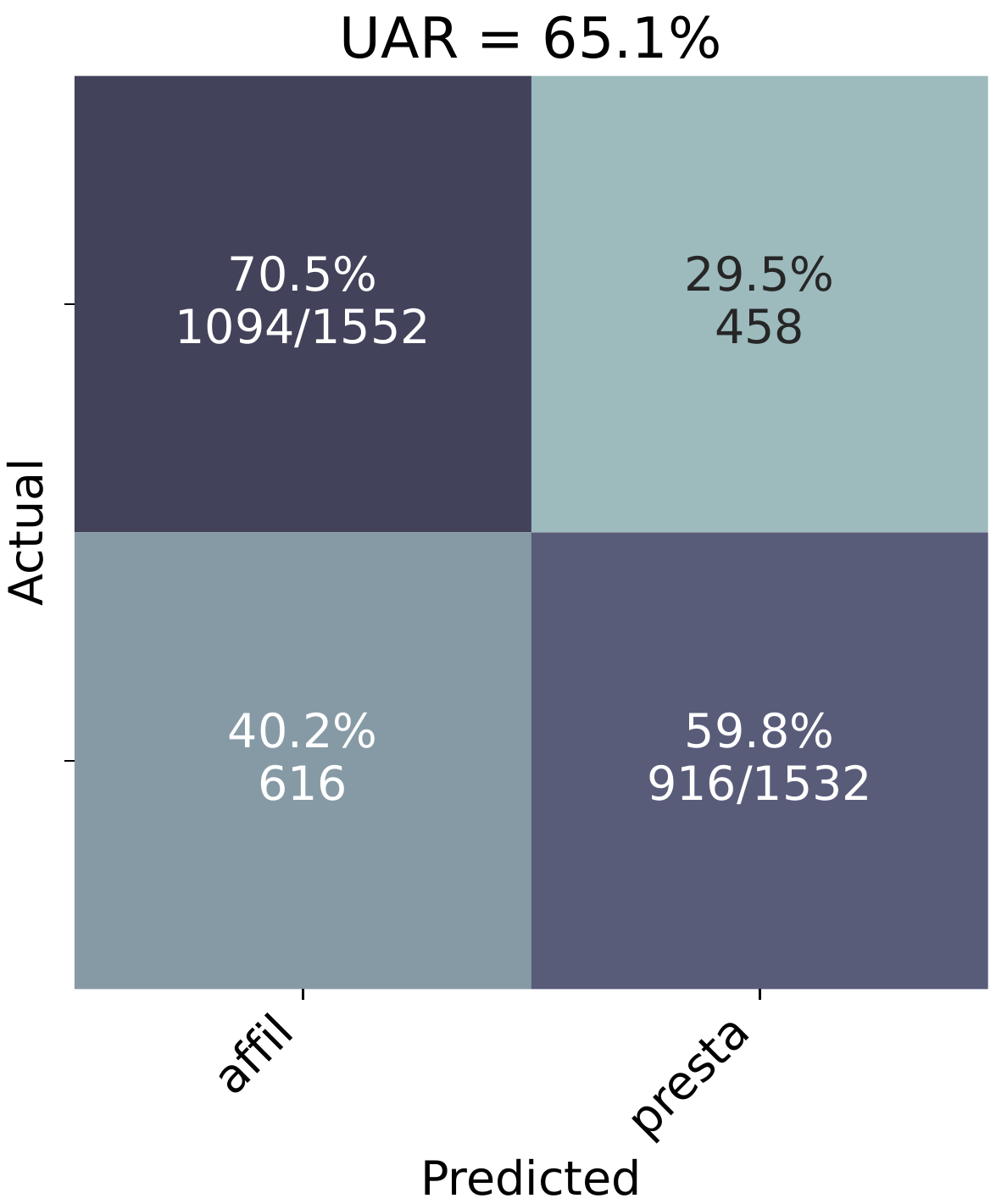}}
    \label{fig:requests-cm}
    \subfloat[HC-C Complaints Confusion Matrix]{
    \includegraphics[width=0.49\columnwidth]{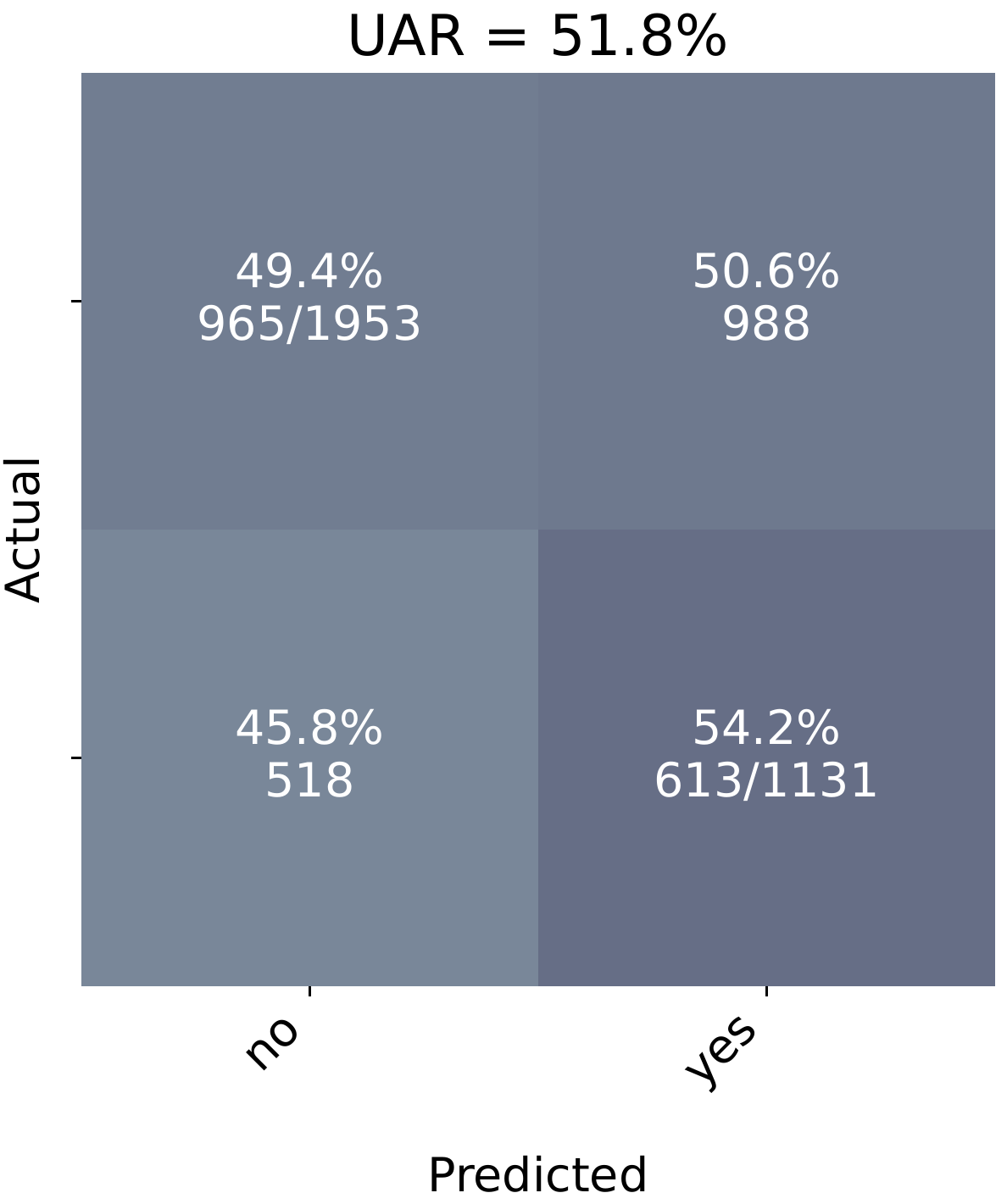}}
    \label{fig:complaints-cm}
\caption{
Confusion matrices for (a) HC-C Requests and (b) HC-C Complaints;
the individual approach/hyperparameters performing best on Dev (without fusion) are chosen; see Table \ref{tab:results}.     
In the cells,   percents of `classified as' of the class displayed in the respective row are given,  also indicated by colour-scale: the darker, the higher. Cases per class  given in  \Cref{tab:db}.}
\label{fig:dev}
\end{figure}

\begin{figure*}[t!]
 \centering
    \subfloat[Anger, $\rho$ $=$ .428]{\includegraphics[width=0.19\linewidth]{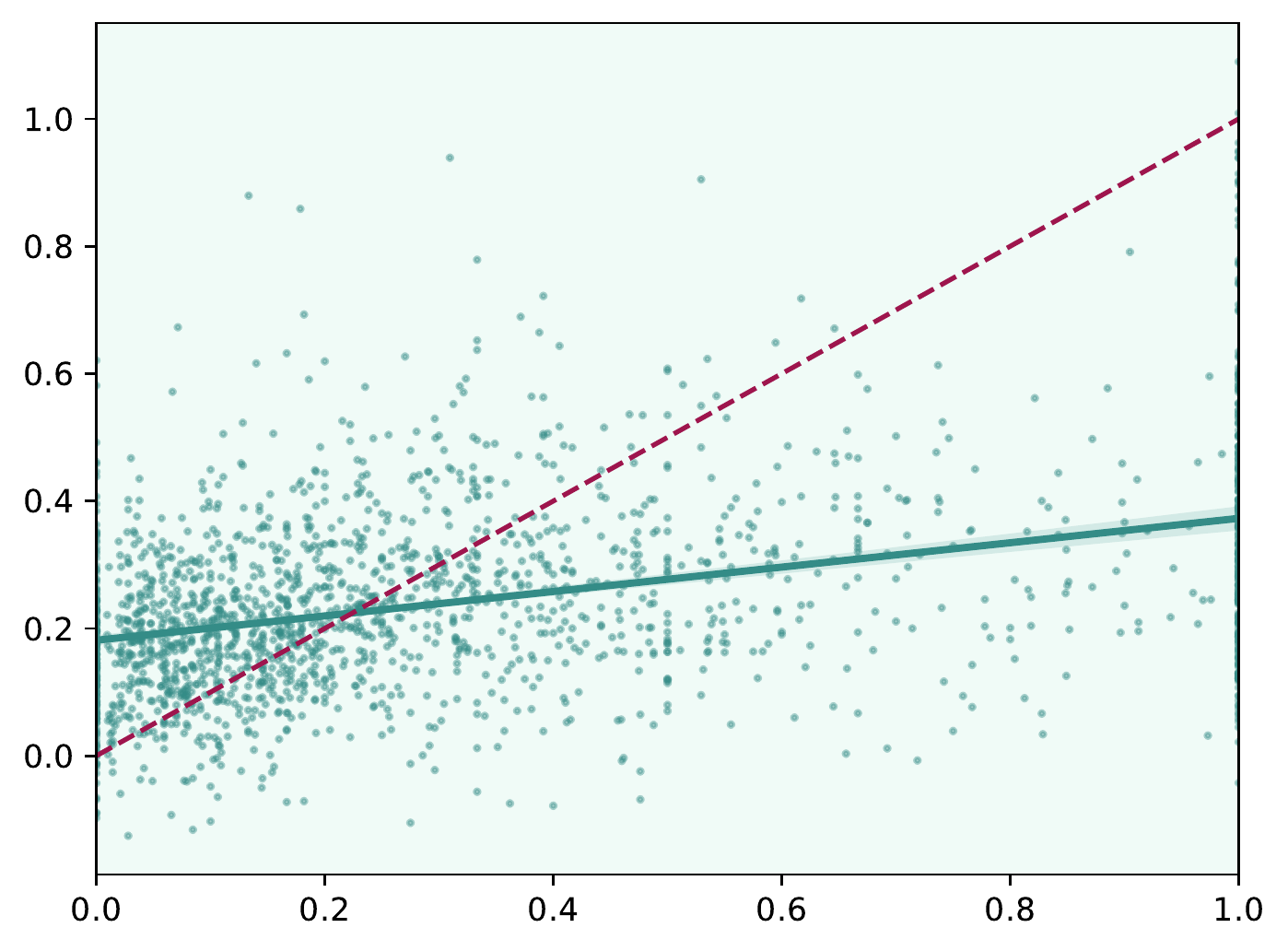}}
    \label{fig:hpc-Anger}
    \hfill
      \subfloat[Interest, $\rho$ $=$ .431]{\includegraphics[width=0.19\linewidth]{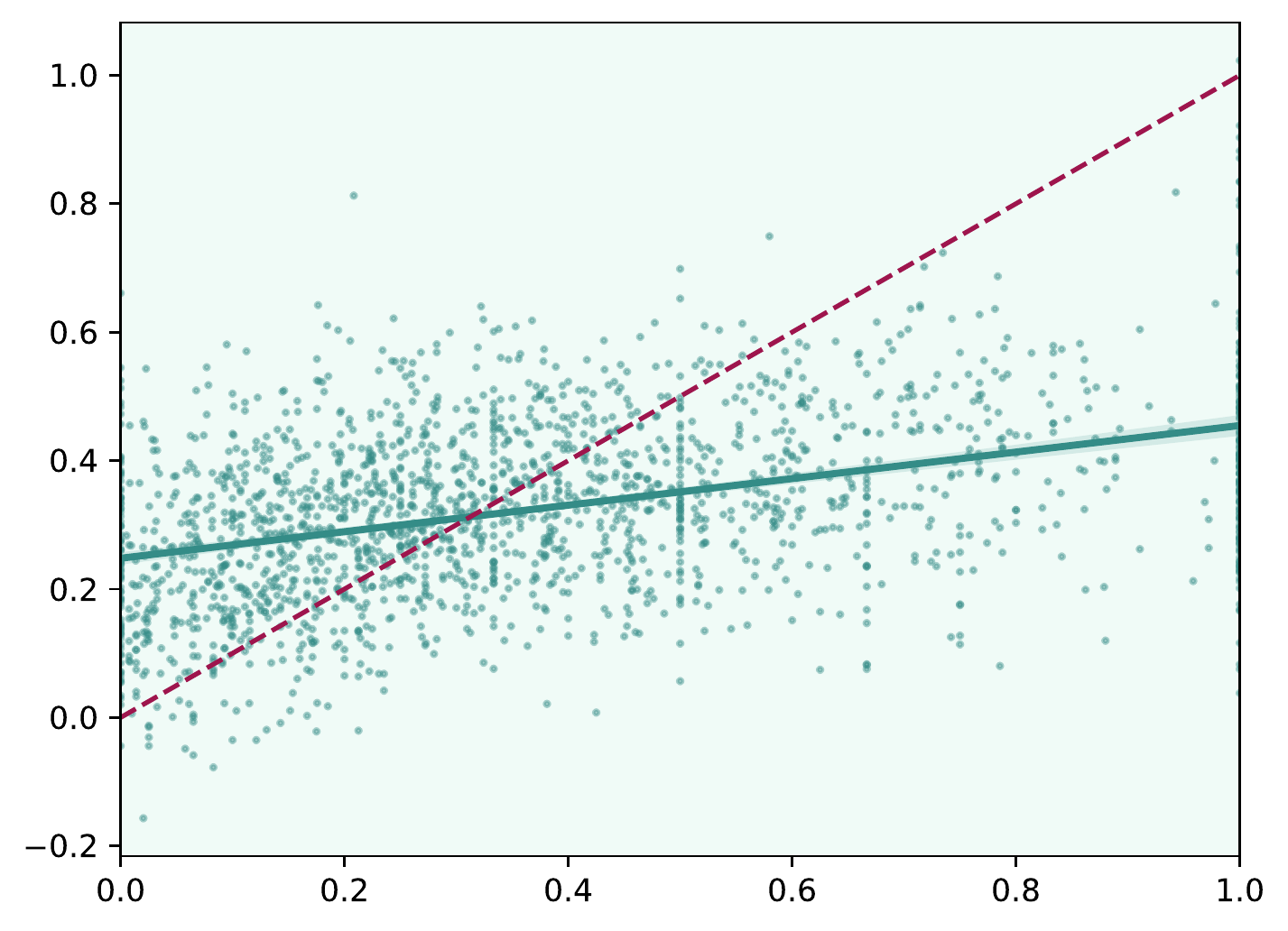}}
    \label{fig:hpc-Interest}
    \hfill
    \subfloat[Excitement, $\rho$ $=$ .453]{\includegraphics[width=0.19\linewidth]{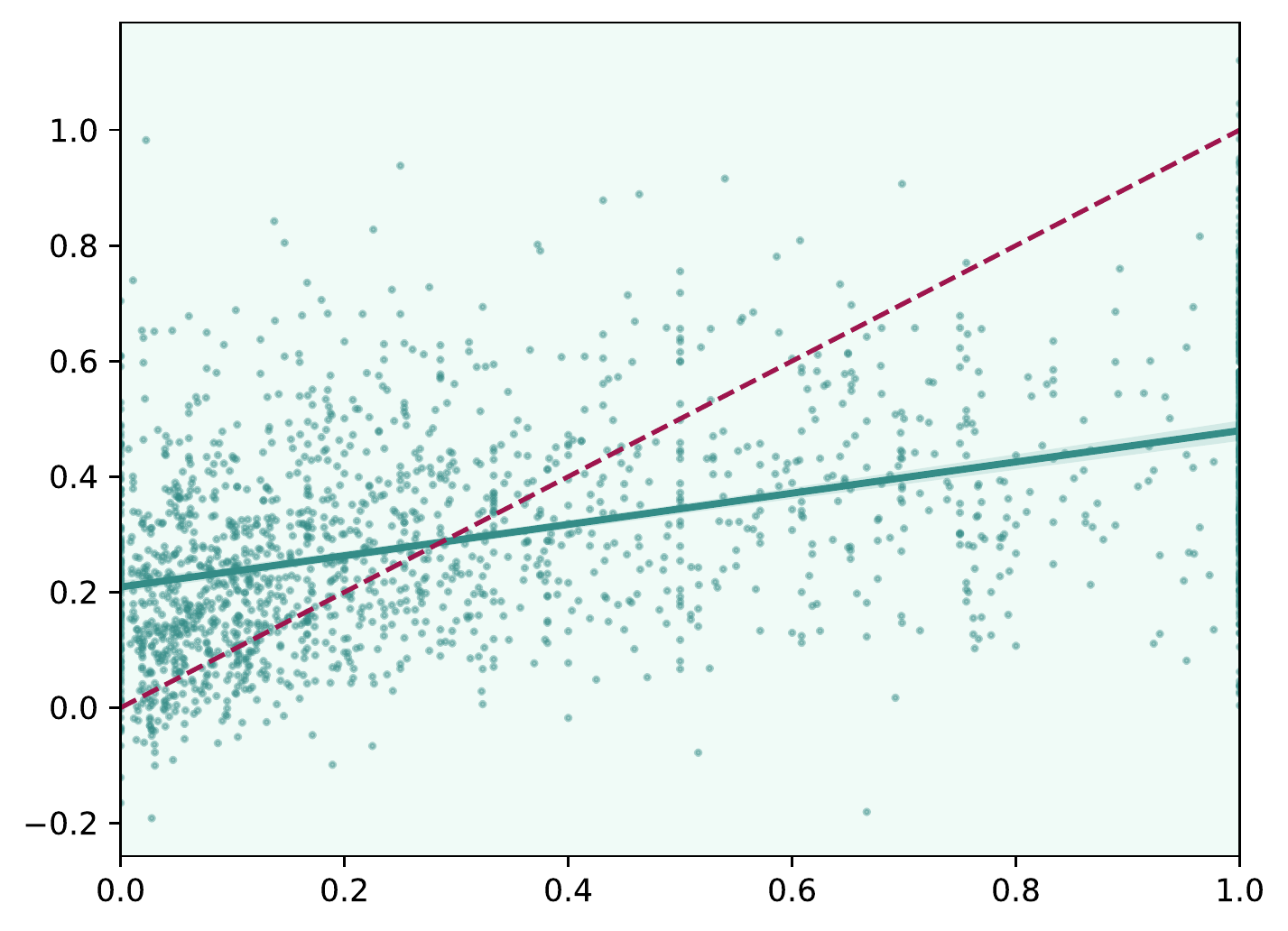}}
    \label{fig:hpc-Excitement}
    \hfill
    \subfloat[Sadness, $\rho$ $=$ .476]{\includegraphics[width=0.19\linewidth]{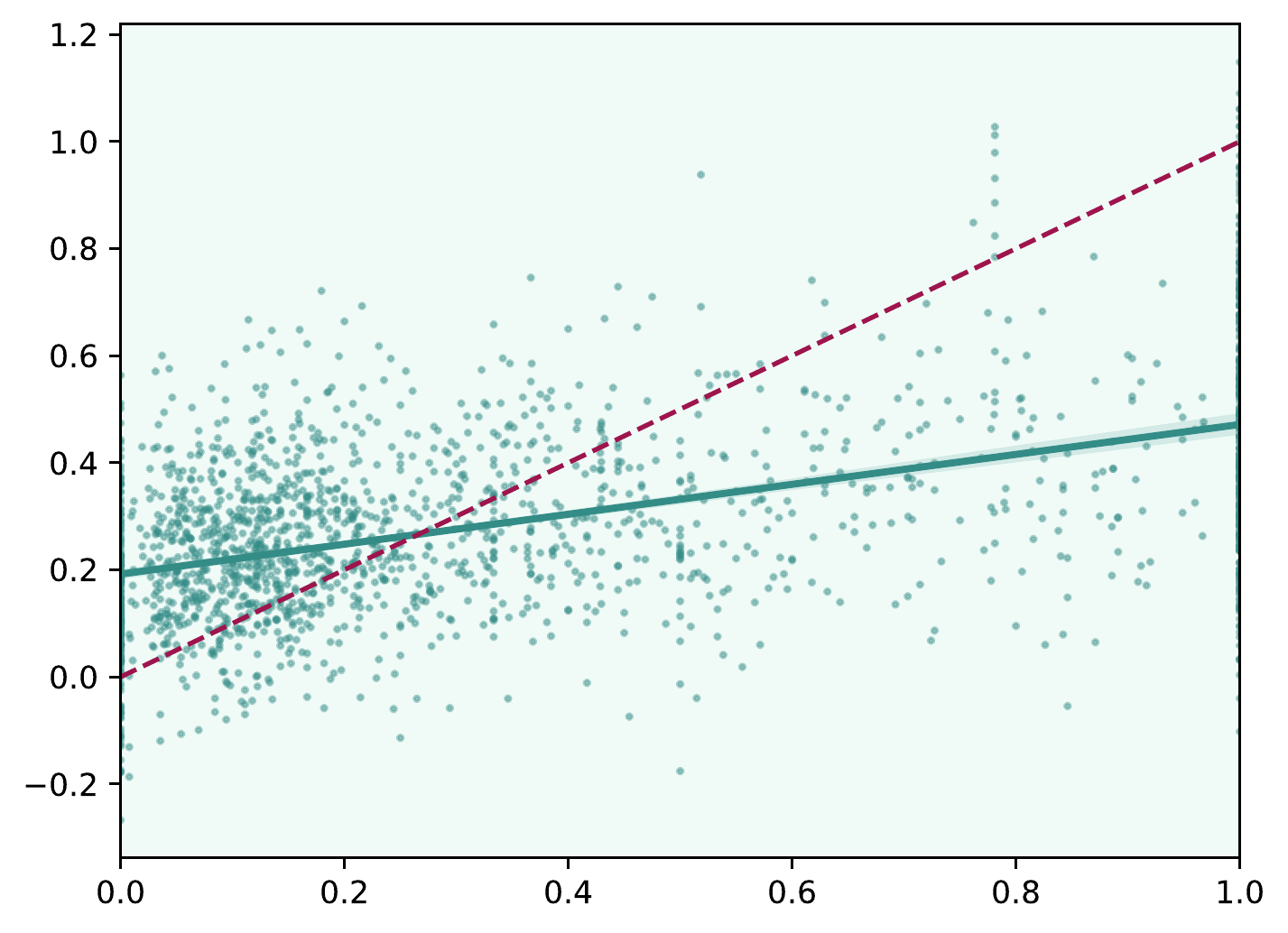}}
    \label{fig:hpc-Sadness}
    \hfill
    
   \subfloat[Concentration, $\rho$ $=$ .524]{\includegraphics[width=0.19\linewidth]{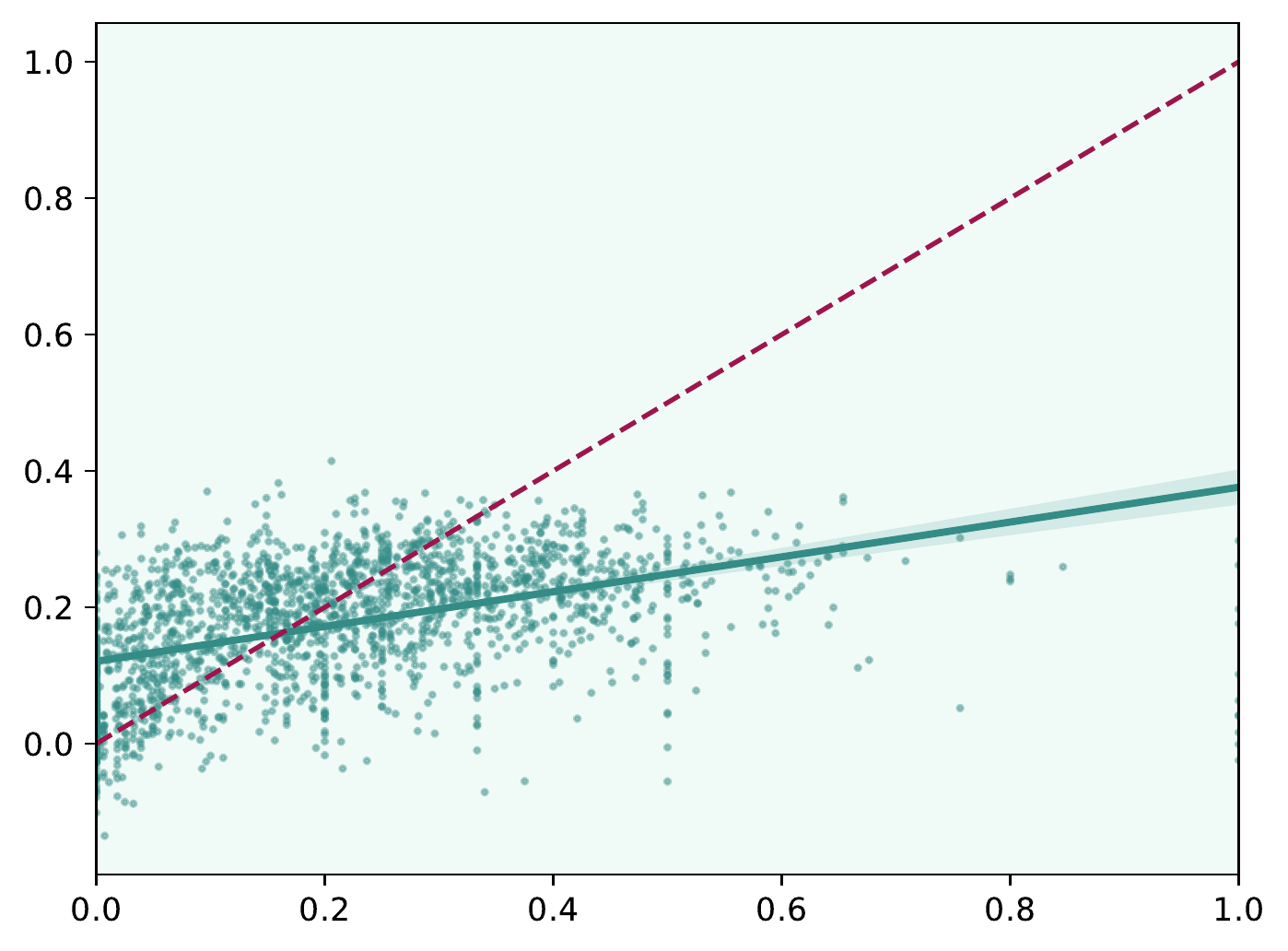}}
    \label{fig:hpc-Concentration}
    \hfill
     \subfloat[Determination, $\rho$ $=$ .531]{\includegraphics[width=0.19\linewidth]{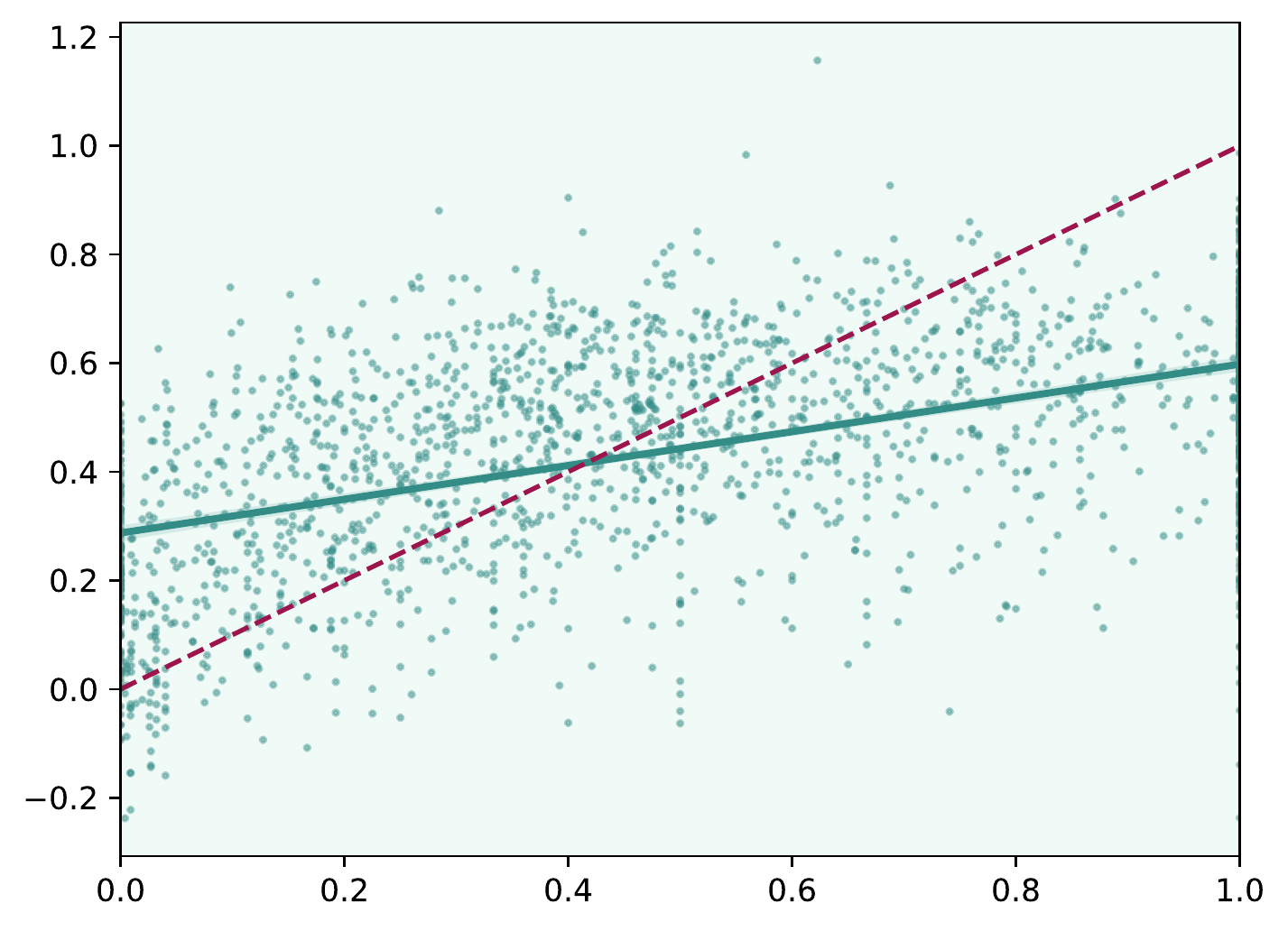}}
    \label{fig:hpc-Determination}
    \hfill
    \subfloat[Boredom, $\rho$ $=$ .545]{\includegraphics[width=0.19\linewidth]{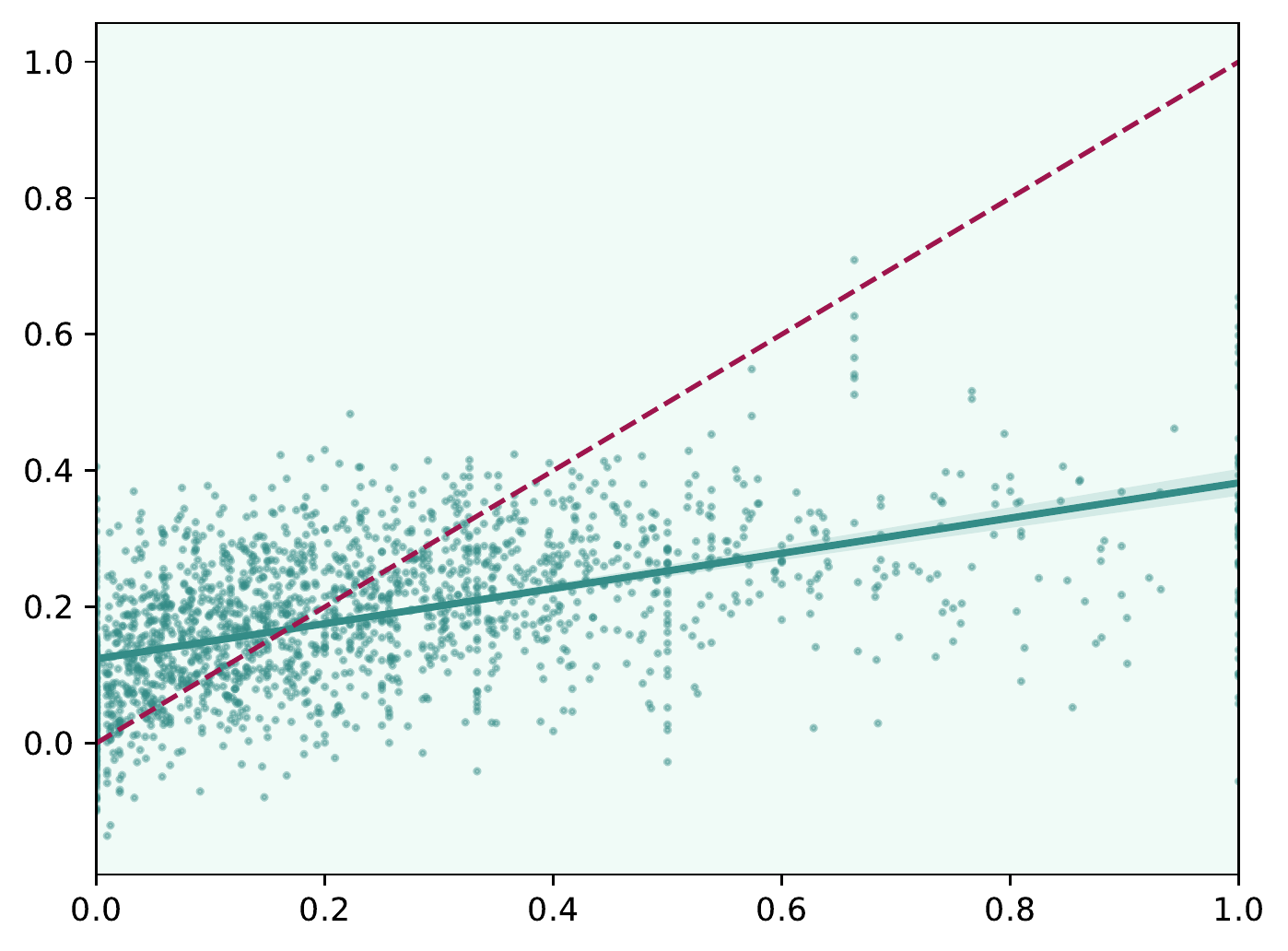}}
    \label{fig:hpc-Boredom}
    \hfill
    \subfloat[Tiredness, $\rho$ $=$ .550]{\includegraphics[width=0.19\linewidth]{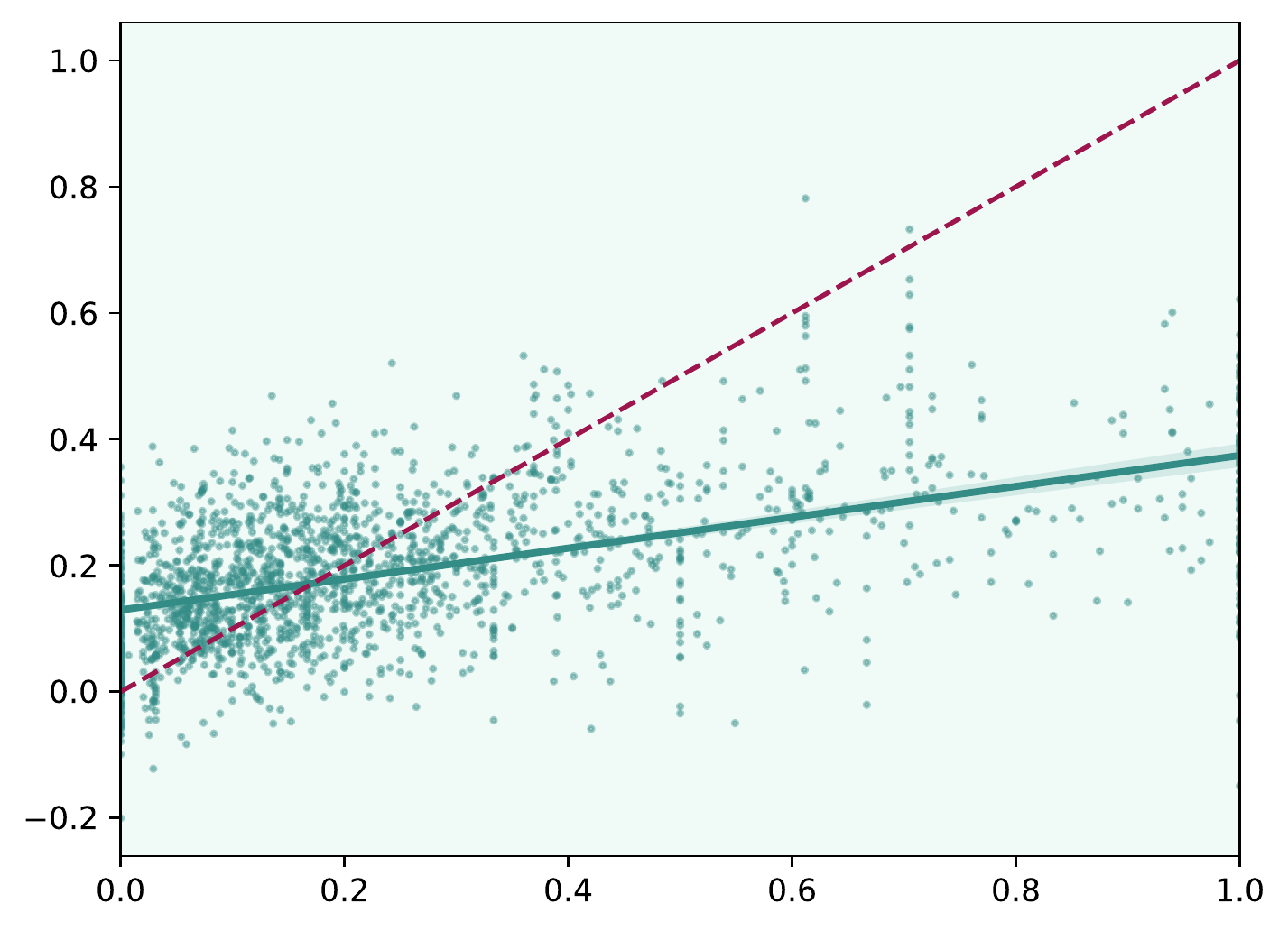}}
    \label{fig:hpc-Tiredness}
    \hfill
        \subfloat[Calmness, $\rho$ $=$ .559]{\includegraphics[width=0.19\linewidth]{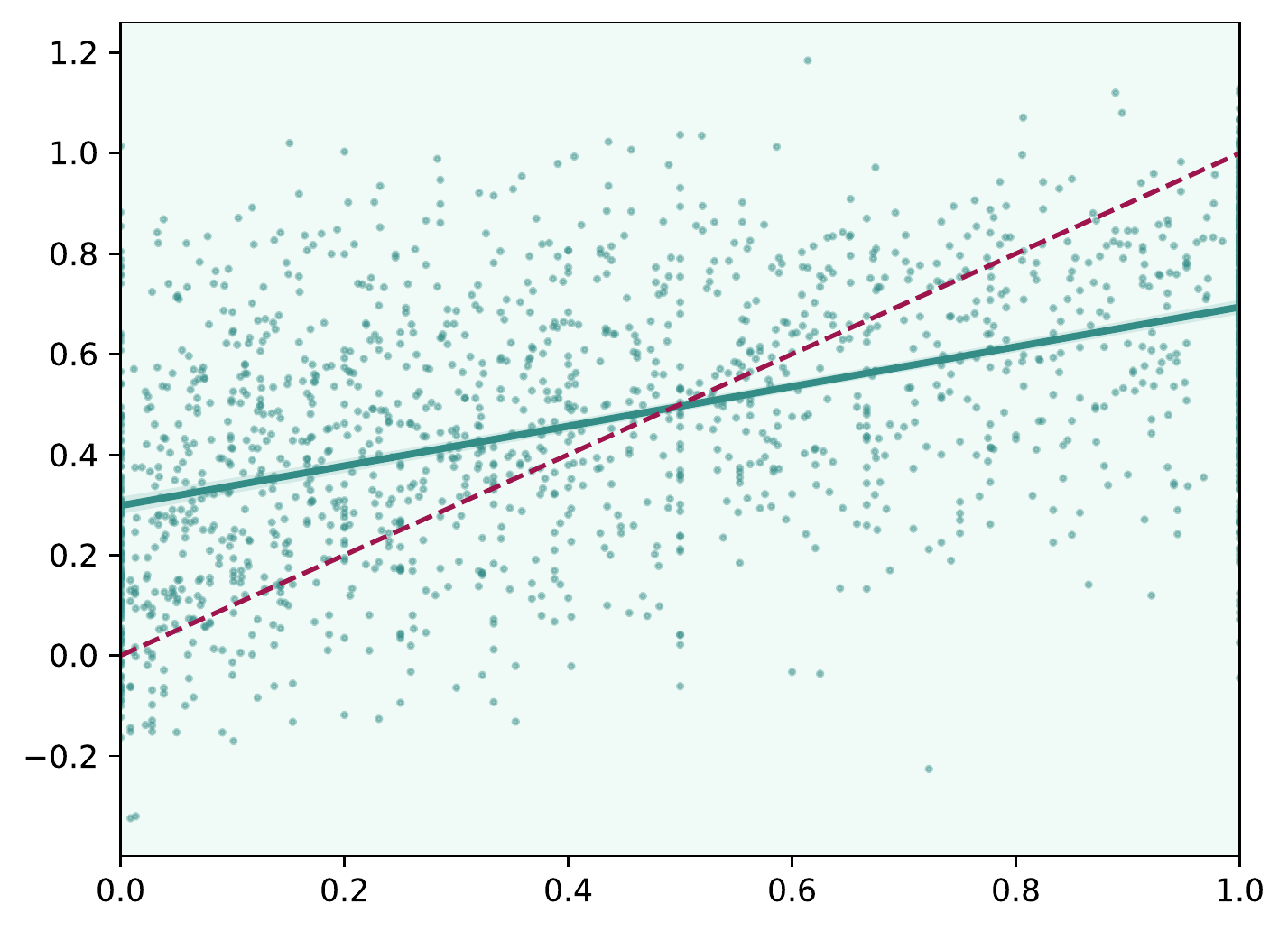}}
    \label{fig:hpc-Calmness}
 \caption{Emotion Share; Scatterplots for each emotion; reference (`true') values on x-axis, predicted values on y-axes; green: regression line; red: `perfect prediction'; random selection of data points for better visibility; emotions ordered by size of $\rho$; for wav2vec2 on Dev, see \Cref{tab:results}.}
\label{fig:hpc}
\end{figure*}

\vspace{-0.1cm}
\subsection{Challenge Baselines and Interpretation}
\label{challenge_baselines}

We provide a branch on the official challenge repository\footnote{\url{https://github.com/EIHW/ComParE2023}} for each Sub-Challenge, which includes scripts allowing participants to fully reproduce the baselines (including pre-processing, model training, and model evaluation on Dev). 
For Requests and Complaints, the 95\,\% Confidence Intervals (CI) were computed by $1\,000$x bootstrapping (random sampling with replacement)  based on the same model that was trained with Train and Dev, and UARs for Test.
For Emotion Share, the 95\,\% Confidence Intervals (CI) were also computed the same way, 
with  Spearmans's correlation score for Test.
(Note that these CIs are too  optimistic because we model single instances which are, however, partly not independent because more than one instance could have  been produced by the same speaker.)


\subsection{The Requests Sub-Challenge}

\noindent
\textbf{Requests:}  
We obtain the best \textbf{UAR=67.2\,\%} on Test with wav2vec2, see  \Cref{tab:results}.  
\Cref{fig:dev}(a) shows, for the best Dev result
given in 
\Cref{tab:results}, that  affil (requests concerning membership issues) can be  better modelled than presta (type of process).
The reason might be that affil  provides less variance in linguistic content than presta, where different types of processes have to be be modelled. 

\noindent
\textbf{Complaints:}
We obtain the best \textbf{UAR=52.9\,\%} on Test with ComParE, see  \Cref{tab:results}.  
\Cref{fig:dev}(b) shows, for the best Dev result given in \Cref{tab:results},
a rather balanced distribution. Yet, UAR is rather low and not really different from chance level. 


Overall baseline for the Requests Sub-Challenge is the combined best UAR of Requests and Complaints: (67.2 + 52.9) / 2 = \textbf{60.1\,\%}.
Note that our baselines are computed with only acoustic information -- concerning wav2vec2, only with the implicit linguistic information entailed in this procedure. Thus, an additional processing of the linguistic content might surely improve performance. 

\subsection{The Emotion Share Sub-Challenge}

We achieve a best UAR on Test of \textbf{$\rho$ = .514} for wav2vec2 which is markedly better than the other three procedures, see \Cref{tab:results}. Note that the data consist of `same' and `different' utterances. As we mentioned above in Section \ref{Corpora}, especially for the `different' sentences,  where the emotional content might at least partly be conveyed with words, the linguistic information entailed in wav2vec2 might contribute to this difference.

\Cref{fig:hpc} displays scatterplots and  regression lines for each emotion, obtained for Dev with wav2vec2. We see that especially calmness and determination are distributed over all reference values. In contrast, especially the prototypical emotions (two of the big~\textit{n}) anger and sadness -- we can attribute excitement to these emotions as well -- are below $\rho$ = .50. This might be explained by the experimental design and the choice of samples which do not favour such prototypical, rather extreme emotions, in contrast to less pronounced ones. The skewed distribution might as well be responsible for the slightly lower performance of the prototypical emotions.

\section{Concluding Remarks}
\label{Conclusion}

This year's challenge is new by two new tasks, all of them highly relevant for applications. 
We feature our `classic' approaches \textbf{ \textsc{ComParE}}, \textbf{ \textsc{auDeep} }, and \textbf{ \ds{}}, and introduce  (fine-tuning of) \textbf{Wav2Vec2} model as an additional baseline.
For all computation steps, scripts are provided that can, but need not be used by the participants.
We expect \mbox{participants} to obtain better performance measures by employing novel (combinations of) procedures and features, including such tailored to the particular tasks.
For both Sub-Challenges -- maybe more for the Requests Sub-Challenge -- additional linguistic modelling might improve performance as well.

\section{Acknowledgments}
We acknowledge funding from the Deutsche Forschungsgemeinschaft  
(DFG) under grant agreement No.\ 421613952 (ParaStiChaD) as well as 
from the DFG's Reinhart Koselleck project No.\ 442218748 
(AUDI0NOMOUS).

\begin{acronym}
\acro{AReLU}[AReLU]{Attention-based Rectified Linear Unit}
\acro{AUC}[AUC]{Area Under the Curve}
\acro{CCC}[CCC]{Concordance Correlation Coefficient}
\acro{CNN}[CNN]{Convolutional Neural Network}
\acrodefplural{CNN}[CNNs]{Convolutional Neural Networks}
\acro{CI}[CI]{Confidence Interval}
\acrodefplural{CI}[CIs]{Confidence Intervals}
\acro{CCS}[CCS]{COVID-19 Cough}
\acro{CSS}[CSS]{COVID-19 Speech}
\acro{CTW}[CTW]{Canonical Time Warping}
\acro{ComParE}[ComParE]{Computational Paralinguistics Challenge}
\acrodefplural{ComParE}[ComParE]{Computational Paralinguistics Challenges}
\acro{DNN}[DNN]{Deep Neural Network}
\acrodefplural{DNNs}[DNNs]{Deep Neural Networks}
\acro{DEMoS}[DEMoS]{Database of Elicited Mood in Speech}
\acro{eGeMAPS}[\textsc{eGeMAPS}]{extended Geneva Minimalistic Acoustic Parameter Set}
\acro{EULA}[EULA]{End User License Agreement}
\acro{EWE}[EWE]{Evaluator Weighted Estimator}
\acro{FLOP}[FLOP]{Floating Point Operation}
\acrodefplural{FLOP}[FLOPs]{Floating Point Operations}
\acro{FAU}[FAU]{Facial Action Unit}
\acrodefplural{FAU}[FAUs]{Facial Action Units}
\acro{GDPR}[GDPR]{General Data Protection Regulation}
\acro{HDF}[HDF]{Hierarchical Data Format}
\acro{Hume-Reaction}[\textsc{Hume-Reaction}]{Hume-Reaction}
\acro{HSQ}[HSQ]{Humor Style Questionnaire}
\acro{IEMOCAP}[IEMOCAP]{Interactive Emotional Dyadic Motion Capture}
\acro{KSS}[KSS]{Karolinska Sleepiness Scale}
\acro{LIME}[LIME]{Local Interpretable Model-agnostic Explanations}
\acro{LLD}[LLD]{Low-Level Descriptor}
\acrodefplural{LLD}[LLDs]{Low-Level Descriptors}
\acro{LSTM}[LSTM]{Long Short-Term Memory}
\acro{MIP}[MIP]{Mood Induction Procedure}
\acro{MIP}[MIPs]{Mood Induction Procedures}
\acro{MLP}[MLP]{Multilayer Perceptron}
\acrodefplural{MLP}[MLPs]{Multilayer Perceptrons}
\acro{MPSSC}[MPSSC]{Munich-Passau Snore Sound Corpus}
\acro{MTCNN}[MTCNN]{Multi-task Cascaded Convolutional Networks}
\acro{SER}[SER]{Speech Emotion Recognition}
\acro{SHAP}[SHAP]{SHapley Additive exPlanations}
\acro{STFT}[STFT]{Short-Time Fourier Transform}
\acrodefplural{STFT}[STFTs]{Short-Time Fourier Transforms}
\acro{SVM}[SVM]{Support Vector Machine}
\acro{TF}[TF]{TensorFlow}
\acro{TSST}[TSST]{Trier Social Stress Test}
\acro{TNR}[TNR]{True Negative Rate}
\acro{TPR}[TPR]{True Positive Rate}
\acro{UAR}[UAR]{Unweighted Average Recall}
\acro{Ulm-TSST}[\textsc{Ulm-TSST}]{Ulm-Trier Social Stress Test}
\acrodefplural{UAR}[UARs]{Unweighted Average Recall}
\end{acronym}

\clearpage
\footnotesize
\bibliographystyle{ACM-Reference-Format}
\balance
\bibliography{sample-base}


\begin{thebibliography}{28}


\ifx \showCODEN    \undefined \def \showCODEN     #1{\unskip}     \fi
\ifx \showDOI      \undefined \def \showDOI       #1{#1}\fi
\ifx \showISBNx    \undefined \def \showISBNx     #1{\unskip}     \fi
\ifx \showISBNxiii \undefined \def \showISBNxiii  #1{\unskip}     \fi
\ifx \showISSN     \undefined \def \showISSN      #1{\unskip}     \fi
\ifx \showLCCN     \undefined \def \showLCCN      #1{\unskip}     \fi
\ifx \shownote     \undefined \def \shownote      #1{#1}          \fi
\ifx \showarticletitle \undefined \def \showarticletitle #1{#1}   \fi
\ifx \showURL      \undefined \def \showURL       {\relax}        \fi
\providecommand\bibfield[2]{#2}
\providecommand\bibinfo[2]{#2}
\providecommand\natexlab[1]{#1}
\providecommand\showeprint[2][]{arXiv:#2}

\bibitem[\protect\citeauthoryear{Amiriparian}{Amiriparian}{2019}]%
        {amiriparian2019deep}
\bibfield{author}{\bibinfo{person}{Shahin Amiriparian}.}
  \bibinfo{year}{2019}\natexlab{}.
\newblock \emph{\bibinfo{title}{Deep Representation Learning Techniques for
  Audio Signal Processing}}.
\newblock \bibinfo{thesistype}{Ph.D. Dissertation}. \bibinfo{school}{Technische
  Universit{\"a}t M{\"u}nchen}.
\newblock


\bibitem[\protect\citeauthoryear{Amiriparian, Freitag, Cummins, and
  Schuller}{Amiriparian et~al\mbox{.}}{2017a}]%
        {Amiriparian17-STS}
\bibfield{author}{\bibinfo{person}{Shahin Amiriparian},
  \bibinfo{person}{Michael Freitag}, \bibinfo{person}{Nicholas Cummins}, {and}
  \bibinfo{person}{Bj\"orn Schuller}.} \bibinfo{year}{2017}\natexlab{a}.
\newblock \showarticletitle{{Sequence to Sequence Autoencoders for Unsupervised
  Representation Learning from Audio}}. In \bibinfo{booktitle}{\emph{Proc.\
  DCASE 2017}}. \bibinfo{publisher}{IEEE}, \bibinfo{address}{Munich, Germany},
  \bibinfo{pages}{17--21}.
\newblock


\bibitem[\protect\citeauthoryear{Amiriparian, Gerczuk, Ottl, Cummins, Freitag,
  Pugachevskiy, and Schuller}{Amiriparian et~al\mbox{.}}{2017b}]%
        {Amiriparian17-SSC}
\bibfield{author}{\bibinfo{person}{Shahin Amiriparian},
  \bibinfo{person}{Maurice Gerczuk}, \bibinfo{person}{Sandra Ottl},
  \bibinfo{person}{Nicholas Cummins}, \bibinfo{person}{Michael Freitag},
  \bibinfo{person}{Sergey Pugachevskiy}, {and} \bibinfo{person}{Bj\"orn
  Schuller}.} \bibinfo{year}{2017}\natexlab{b}.
\newblock \showarticletitle{Snore Sound Classification Using Image-based Deep
  Spectrum Features}. In \bibinfo{booktitle}{\emph{Proc.\ Interspeech}}.
  \bibinfo{publisher}{ISCA}, \bibinfo{address}{Stockholm, Sweden},
  \bibinfo{pages}{3512--3516}.
\newblock


\bibitem[\protect\citeauthoryear{Amiriparian, Gerczuk, Stappen, Baird, Koebe,
  Ottl, and Schuller}{Amiriparian et~al\mbox{.}}{2020}]%
        {Amiriparian20-TCP}
\bibfield{author}{\bibinfo{person}{Shahin Amiriparian},
  \bibinfo{person}{Maurice Gerczuk}, \bibinfo{person}{Lukas Stappen},
  \bibinfo{person}{Alice Baird}, \bibinfo{person}{Lukas Koebe},
  \bibinfo{person}{Sandra Ottl}, {and} \bibinfo{person}{Bj\"orn Schuller}.}
  \bibinfo{year}{2020}\natexlab{}.
\newblock \showarticletitle{{Towards Cross-Modal Pre-Training and Learning
  Tempo-Spatial Characteristics for Audio Recognition with Convolutional and
  Recurrent Neural Networks}}.
\newblock \bibinfo{journal}{\emph{EURASIP Journal on Audio, Speech, and Music
  Processing}} \bibinfo{volume}{2020}, \bibinfo{number}{19}
  (\bibinfo{year}{2020}), \bibinfo{pages}{1--11}.
\newblock


\bibitem[\protect\citeauthoryear{Anshari, Almunawar, Lim, and
  Al-Mudimigh}{Anshari et~al\mbox{.}}{2019}]%
        {Anshari19-CRM}
\bibfield{author}{\bibinfo{person}{Muhammad Anshari},
  \bibinfo{person}{Mohammad~Nabil Almunawar}, \bibinfo{person}{Syamimi~Ariff
  Lim}, {and} \bibinfo{person}{Abdullah Al-Mudimigh}.}
  \bibinfo{year}{2019}\natexlab{}.
\newblock \showarticletitle{{Customer relationship management and big data
  enabled: Personalization \& customization of services}}.
\newblock \bibinfo{journal}{\emph{Applied Computing and Informatics}}
  \bibinfo{volume}{15} (\bibinfo{year}{2019}), \bibinfo{pages}{94--101}.
\newblock


\bibitem[\protect\citeauthoryear{Batliner, Hantke, and Schuller}{Batliner
  et~al\mbox{.}}{2020}]%
        {batliner2020ethics}
\bibfield{author}{\bibinfo{person}{Anton Batliner}, \bibinfo{person}{Simone
  Hantke}, {and} \bibinfo{person}{Bj{\"o}rn Schuller}.}
  \bibinfo{year}{2020}\natexlab{}.
\newblock \showarticletitle{Ethics and good practice in computational
  paralinguistics}.
\newblock \bibinfo{journal}{\emph{IEEE Transactions on Affective Computing}}
  \bibinfo{volume}{13}, \bibinfo{number}{3} (\bibinfo{year}{2020}),
  \bibinfo{pages}{1236--1253}.
\newblock


\bibitem[\protect\citeauthoryear{Cowen, Elfenbein, Laukka, and Keltner}{Cowen
  et~al\mbox{.}}{2019a}]%
        {Cowen19-M2E}
\bibfield{author}{\bibinfo{person}{Alan~S. Cowen},
  \bibinfo{person}{Hillary~Anger Elfenbein}, \bibinfo{person}{Petri Laukka},
  {and} \bibinfo{person}{Dacher Keltner}.} \bibinfo{year}{2019}\natexlab{a}.
\newblock \showarticletitle{{Mapping 24 emotions conveyed by brief human
  vocalization}}.
\newblock \bibinfo{journal}{\emph{Am Psychol.}}  \bibinfo{volume}{74}
  (\bibinfo{year}{2019}), \bibinfo{pages}{698--712}.
\newblock


\bibitem[\protect\citeauthoryear{Cowen and Keltner}{Cowen and Keltner}{2021}]%
        {cowen2021semantic}
\bibfield{author}{\bibinfo{person}{Alan~S Cowen} {and} \bibinfo{person}{Dacher
  Keltner}.} \bibinfo{year}{2021}\natexlab{}.
\newblock \showarticletitle{Semantic space theory: A computational approach to
  emotion}.
\newblock \bibinfo{journal}{\emph{Trends in Cognitive Sciences}}
  \bibinfo{volume}{25}, \bibinfo{number}{2} (\bibinfo{year}{2021}),
  \bibinfo{pages}{124--136}.
\newblock


\bibitem[\protect\citeauthoryear{Cowen, Laukka, Elfenbein, Liu, and
  Keltner}{Cowen et~al\mbox{.}}{2019b}]%
        {cowen2019primacy}
\bibfield{author}{\bibinfo{person}{Alan~S Cowen}, \bibinfo{person}{Petri
  Laukka}, \bibinfo{person}{Hillary~Anger Elfenbein}, \bibinfo{person}{Runjing
  Liu}, {and} \bibinfo{person}{Dacher Keltner}.}
  \bibinfo{year}{2019}\natexlab{b}.
\newblock \showarticletitle{The primacy of categories in the recognition of 12
  emotions in speech prosody across two cultures}.
\newblock \bibinfo{journal}{\emph{Nature human behaviour}} \bibinfo{volume}{3},
  \bibinfo{number}{4} (\bibinfo{year}{2019}), \bibinfo{pages}{369--382}.
\newblock


\bibitem[\protect\citeauthoryear{Elfenbein, Laukka, Althoff, Chui, Iraki,
  Rockstuhl, and Thingujam}{Elfenbein et~al\mbox{.}}{2022}]%
        {elfenbein2022we}
\bibfield{author}{\bibinfo{person}{Hillary~Anger Elfenbein},
  \bibinfo{person}{Petri Laukka}, \bibinfo{person}{Jean Althoff},
  \bibinfo{person}{Wanda Chui}, \bibinfo{person}{Frederick~K Iraki},
  \bibinfo{person}{Thomas Rockstuhl}, {and} \bibinfo{person}{Nutankumar~S.
  Thingujam}.} \bibinfo{year}{2022}\natexlab{}.
\newblock \showarticletitle{What Do We Hear in the Voice? An Open-Ended
  Judgment Study of Emotional Speech Prosody}.
\newblock \bibinfo{journal}{\emph{Personality and Social Psychology Bulletin}}
  \bibinfo{volume}{48}, \bibinfo{number}{7} (\bibinfo{year}{2022}),
  \bibinfo{pages}{1087--1104}.
\newblock


\bibitem[\protect\citeauthoryear{Eyben, Weninger, Gro{\ss}, and Schuller}{Eyben
  et~al\mbox{.}}{2013}]%
        {Eyben13-RDI}
\bibfield{author}{\bibinfo{person}{Florian Eyben}, \bibinfo{person}{Felix
  Weninger}, \bibinfo{person}{Florian Gro{\ss}}, {and} \bibinfo{person}{Bj\"orn
  Schuller}.} \bibinfo{year}{2013}\natexlab{}.
\newblock \showarticletitle{{Recent Developments in {openSMILE}, the {M}unich
  Open-Source Multimedia Feature Extractor}}. In
  \bibinfo{booktitle}{\emph{{Proc.\ ACM Multimedia}}}.
  \bibinfo{publisher}{ACM}, \bibinfo{address}{Barcelona, Spain},
  \bibinfo{pages}{835--838}.
\newblock


\bibitem[\protect\citeauthoryear{Freitag, Amiriparian, Pugachevskiy, Cummins,
  and Schuller}{Freitag et~al\mbox{.}}{2018}]%
        {Freitag18-AUL}
\bibfield{author}{\bibinfo{person}{Michael Freitag}, \bibinfo{person}{Shahin
  Amiriparian}, \bibinfo{person}{Sergey Pugachevskiy},
  \bibinfo{person}{Nicholas Cummins}, {and} \bibinfo{person}{Bj\"orn
  Schuller}.} \bibinfo{year}{2018}\natexlab{}.
\newblock \showarticletitle{{{auDeep}: Unsupervised Learning of Representations
  from Audio with Deep Recurrent Neural Networks}}.
\newblock \bibinfo{journal}{\emph{Journal of Machine Learning Research}}
  \bibinfo{volume}{18} (\bibinfo{year}{2018}), \bibinfo{pages}{1--5}.
\newblock


\bibitem[\protect\citeauthoryear{Hildebrand, Efthymiou, Busquet, Hampton,
  Hoffman, and Novak}{Hildebrand et~al\mbox{.}}{2020}]%
        {Hildebrand20-VAI}
\bibfield{author}{\bibinfo{person}{Christian Hildebrand},
  \bibinfo{person}{Fotis Efthymiou}, \bibinfo{person}{Francesc Busquet},
  \bibinfo{person}{William~H. Hampton}, \bibinfo{person}{Donna~L. Hoffman},
  {and} \bibinfo{person}{Thomas~P. Novak}.} \bibinfo{year}{2020}\natexlab{}.
\newblock \showarticletitle{{Voice analytics in business research. Conceptual
  foundations, acoustic feature extraction, and applications}}.
\newblock \bibinfo{journal}{\emph{Journal of Business Research}}
  \bibinfo{volume}{121} (\bibinfo{year}{2020}), \bibinfo{pages}{364--374}.
\newblock


\bibitem[\protect\citeauthoryear{Lackovic, Montacié, Lalande, and
  Caraty}{Lackovic et~al\mbox{.}}{2022}]%
        {Lackovic22-POU}
\bibfield{author}{\bibinfo{person}{Nikola Lackovic}, \bibinfo{person}{Claude
  Montacié}, \bibinfo{person}{Gauthier Lalande}, {and}
  \bibinfo{person}{Marie-José Caraty}.} \bibinfo{year}{2022}\natexlab{}.
\newblock \showarticletitle{Prediction of User Request and Complaint in Spoken
  Customer-Agent Conversations}.
\newblock \bibinfo{journal}{\emph{preprint arXiv:2208.10249}}
  (\bibinfo{year}{2022}).
\newblock


\bibitem[\protect\citeauthoryear{Laukka, Elfenbein, Chui, Thingujam, Iraki,
  Rockstuhl, and Althoff}{Laukka et~al\mbox{.}}{2010}]%
        {Laukka10-PTV}
\bibfield{author}{\bibinfo{person}{Petri Laukka},
  \bibinfo{person}{Hillary~Anger Elfenbein}, \bibinfo{person}{Wanda Chui},
  \bibinfo{person}{Nutankumar~S. Thingujam}, \bibinfo{person}{Frederick~K.
  Iraki}, \bibinfo{person}{Thomas Rockstuhl}, {and} \bibinfo{person}{Jean
  Althoff}.} \bibinfo{year}{2010}\natexlab{}.
\newblock \showarticletitle{{Presenting the VENEC corpus: Development of a
  cross-cultural corpus of vocal emotion expressions and a novel method of
  annotating emotion appraisals}}.
\newblock In \bibinfo{booktitle}{\emph{Proc.\ LREC 2010 Workshop on Corpora for
  Research on Emotion and Affect}}. \bibinfo{publisher}{LREC},
  \bibinfo{address}{Marrakesh, Morocco}, \bibinfo{pages}{53--57}.
\newblock


\bibitem[\protect\citeauthoryear{Laukka, Elfenbein, Thingujam, Rockstuhl,
  Iraki, Chui, and Althoff}{Laukka et~al\mbox{.}}{2016}]%
        {laukka2016expression}
\bibfield{author}{\bibinfo{person}{Petri Laukka},
  \bibinfo{person}{Hillary~Anger Elfenbein}, \bibinfo{person}{Nutankumar~S
  Thingujam}, \bibinfo{person}{Thomas Rockstuhl}, \bibinfo{person}{Frederick~K
  Iraki}, \bibinfo{person}{Wanda Chui}, {and} \bibinfo{person}{Jean Althoff}.}
  \bibinfo{year}{2016}\natexlab{}.
\newblock \showarticletitle{The expression and recognition of emotions in the
  voice across five nations: A lens model analysis based on acoustic features.}
\newblock \bibinfo{journal}{\emph{Journal of personality and social
  psychology}} \bibinfo{volume}{111}, \bibinfo{number}{5}
  (\bibinfo{year}{2016}), \bibinfo{pages}{686}.
\newblock


\bibitem[\protect\citeauthoryear{Lotfian and Busso}{Lotfian and Busso}{2019}]%
        {Lotfian_2019_3}
\bibfield{author}{\bibinfo{person}{Reza Lotfian} {and} \bibinfo{person}{Carlos
  Busso}.} \bibinfo{year}{2019}\natexlab{}.
\newblock \showarticletitle{Building Naturalistic Emotionally Balanced Speech
  Corpus by Retrieving Emotional Speech From Existing Podcast Recordings}.
\newblock \bibinfo{journal}{\emph{IEEE Transactions on Affective Computing}}
  \bibinfo{volume}{10}, \bibinfo{number}{4} (\bibinfo{year}{2019}),
  \bibinfo{pages}{471--483}.
\newblock


\bibitem[\protect\citeauthoryear{Poria, Hazarika, Majumder, Naik, Cambria, and
  Mihalcea}{Poria et~al\mbox{.}}{2018}]%
        {poria2018meld}
\bibfield{author}{\bibinfo{person}{Soujanya Poria}, \bibinfo{person}{Devamanyu
  Hazarika}, \bibinfo{person}{Navonil Majumder}, \bibinfo{person}{Gautam Naik},
  \bibinfo{person}{Erik Cambria}, {and} \bibinfo{person}{Rada Mihalcea}.}
  \bibinfo{year}{2018}\natexlab{}.
\newblock \showarticletitle{Meld: A multimodal multi-party dataset for emotion
  recognition in conversations}.
\newblock \bibinfo{journal}{\emph{preprint arXiv:1810.02508}}
  (\bibinfo{year}{2018}).
\newblock


\bibitem[\protect\citeauthoryear{Rosenberg}{Rosenberg}{2012}]%
        {Rosenberg12-CSD}
\bibfield{author}{\bibinfo{person}{Andrew Rosenberg}.}
  \bibinfo{year}{2012}\natexlab{}.
\newblock \showarticletitle{{Classifying skewed data: importance weighting to
  optimize average recall}}. In \bibinfo{booktitle}{\emph{Proc.\ Interspeech}}.
  \bibinfo{publisher}{ISCA}, \bibinfo{address}{Portland, USA},
  \bibinfo{pages}{2242--2245}.
\newblock


\bibitem[\protect\citeauthoryear{Scheidt and Chung}{Scheidt and Chung}{2019}]%
        {Scheidt19-MAC}
\bibfield{author}{\bibinfo{person}{Scott Scheidt} {and} \bibinfo{person}{Q.~B.
  Chung}.} \bibinfo{year}{2019}\natexlab{}.
\newblock \showarticletitle{{Making a case for speech analytics to improve
  customer service quality: Vision, implementation and evaluation}}.
\newblock \bibinfo{journal}{\emph{International Journal of Information
  Management}}  \bibinfo{volume}{45} (\bibinfo{year}{2019}),
  \bibinfo{pages}{223--232}.
\newblock


\bibitem[\protect\citeauthoryear{Schuller and Batliner}{Schuller and
  Batliner}{2014}]%
        {Schuller14-CPE}
\bibfield{author}{\bibinfo{person}{Björn Schuller} {and}
  \bibinfo{person}{Anton Batliner}.} \bibinfo{year}{2014}\natexlab{}.
\newblock \bibinfo{booktitle}{\emph{Computational Paralinguistics -- Emotion,
  Affect, and Personality in Speech and Language Processing}}.
\newblock \bibinfo{publisher}{Wiley}, \bibinfo{address}{Chichester, UK}.
\newblock


\bibitem[\protect\citeauthoryear{Schuller, Batliner, Steidl, and
  Seppi}{Schuller et~al\mbox{.}}{2011}]%
        {Schuller11-RRE}
\bibfield{author}{\bibinfo{person}{Bj\"orn Schuller}, \bibinfo{person}{Anton
  Batliner}, \bibinfo{person}{Stefan Steidl}, {and} \bibinfo{person}{Dino
  Seppi}.} \bibinfo{year}{2011}\natexlab{}.
\newblock \showarticletitle{{Recognising Realistic Emotions and Affect in
  Speech: State of the Art and Lessons Learnt from the First {Challenge}}}.
\newblock \bibinfo{journal}{\emph{Speech Communication}}  \bibinfo{volume}{53}
  (\bibinfo{year}{2011}), \bibinfo{pages}{1062--1087}.
\newblock


\bibitem[\protect\citeauthoryear{Schuller, Steidl, and Batliner}{Schuller
  et~al\mbox{.}}{2009}]%
        {Schuller09-TI2}
\bibfield{author}{\bibinfo{person}{Björn Schuller}, \bibinfo{person}{Stefan
  Steidl}, {and} \bibinfo{person}{Anton Batliner}.}
  \bibinfo{year}{2009}\natexlab{}.
\newblock \showarticletitle{{The INTERSPEECH 2009 Emotion Challenge}}. In
  \bibinfo{booktitle}{\emph{{Proc.\ Interspeech}}}. \bibinfo{publisher}{ISCA},
  \bibinfo{address}{Brighton, UK}, \bibinfo{pages}{312--315}.
\newblock


\bibitem[\protect\citeauthoryear{Schuller, Steidl, Batliner, Vinciarelli,
  Scherer, Ringeval, Chetouani, Weninger, Eyben, Marchi, Mortillaro, Salamin,
  Polychroniou, Valente, and Kim}{Schuller et~al\mbox{.}}{2013}]%
        {Schuller13-TI2}
\bibfield{author}{\bibinfo{person}{Bj\"orn Schuller}, \bibinfo{person}{Stefan
  Steidl}, \bibinfo{person}{Anton Batliner}, \bibinfo{person}{Alessandro
  Vinciarelli}, \bibinfo{person}{Klaus Scherer}, \bibinfo{person}{Fabien
  Ringeval}, \bibinfo{person}{Mohamed Chetouani}, \bibinfo{person}{Felix
  Weninger}, \bibinfo{person}{Florian Eyben}, \bibinfo{person}{Erik Marchi},
  \bibinfo{person}{Marcello Mortillaro}, \bibinfo{person}{Hugues Salamin},
  \bibinfo{person}{Anna Polychroniou}, \bibinfo{person}{Fabio Valente}, {and}
  \bibinfo{person}{Samuel Kim}.} \bibinfo{year}{2013}\natexlab{}.
\newblock \showarticletitle{{The INTERSPEECH 2013 Computational Paralinguistics
  Challenge: Social Signals, Conflict, Emotion, Autism}}. In
  \bibinfo{booktitle}{\emph{Proc.\ Interspeech}}. \bibinfo{publisher}{ISCA},
  \bibinfo{address}{Lyon, France}, \bibinfo{pages}{148--152}.
\newblock


\bibitem[\protect\citeauthoryear{Schuller, Batliner, Bergler, Mascolo, Han,
  Lefter, Kaya, Amiriparian, Baird, Stappen, Ottl, Gerczuk, Tzirakis, Brown,
  Chauhan, Grammenos, Hasthanasombat, Spathis, Xia, Cicuta, Rothkrantz, Zwerts,
  Treep, and Kaandorp}{Schuller et~al\mbox{.}}{2021}]%
        {schuller2021interspeech}
\bibfield{author}{\bibinfo{person}{Björn~W. Schuller}, \bibinfo{person}{Anton
  Batliner}, \bibinfo{person}{Christian Bergler}, \bibinfo{person}{Cecilia
  Mascolo}, \bibinfo{person}{Jing Han}, \bibinfo{person}{Iulia Lefter},
  \bibinfo{person}{Heysem Kaya}, \bibinfo{person}{Shahin Amiriparian},
  \bibinfo{person}{Alice Baird}, \bibinfo{person}{Lukas Stappen},
  \bibinfo{person}{Sandra Ottl}, \bibinfo{person}{Maurice Gerczuk},
  \bibinfo{person}{Panagiotis Tzirakis}, \bibinfo{person}{Chloë Brown},
  \bibinfo{person}{Jagmohan Chauhan}, \bibinfo{person}{Andreas Grammenos},
  \bibinfo{person}{Apinan Hasthanasombat}, \bibinfo{person}{Dimitris Spathis},
  \bibinfo{person}{Tong Xia}, \bibinfo{person}{Pietro Cicuta},
  \bibinfo{person}{Leon J.~M. Rothkrantz}, \bibinfo{person}{Joeri Zwerts},
  \bibinfo{person}{Jelle Treep}, {and} \bibinfo{person}{Casper Kaandorp}.}
  \bibinfo{year}{2021}\natexlab{}.
\newblock \showarticletitle{The INTERSPEECH 2021 Computational Paralinguistics
  Challenge: COVID-19 Cough, COVID-19 Speech, Escalation \& Primates}. In
  \bibinfo{booktitle}{\emph{Proc.\ Interspeech}}. \bibinfo{publisher}{ISCA},
  \bibinfo{address}{Brno, Czechia}, \bibinfo{pages}{431--435}.
\newblock


\bibitem[\protect\citeauthoryear{Schuller, Batliner, Bergler, Messner,
  Hamilton, Amiriparian, Baird, Rizos, Schmitt, Stappen,
  et~al\mbox{.}}{Schuller et~al\mbox{.}}{2020}]%
        {schuller2020interspeech}
\bibfield{author}{\bibinfo{person}{Bj{\"o}rn~W Schuller},
  \bibinfo{person}{Anton Batliner}, \bibinfo{person}{Christian Bergler},
  \bibinfo{person}{Eva-Maria Messner}, \bibinfo{person}{Antonia Hamilton},
  \bibinfo{person}{Shahin Amiriparian}, \bibinfo{person}{Alice Baird},
  \bibinfo{person}{Georgios Rizos}, \bibinfo{person}{Maximilian Schmitt},
  \bibinfo{person}{Lukas Stappen}, {et~al\mbox{.}}}
  \bibinfo{year}{2020}\natexlab{}.
\newblock \showarticletitle{The INTERSPEECH 2020 Computational Paralinguistics
  Challenge: Elderly Emotion, Breathing \& Masks}. In
  \bibinfo{booktitle}{\emph{Proc.\ Interspeech}}. \bibinfo{publisher}{ISCA},
  \bibinfo{address}{Shanghai, China}, \bibinfo{pages}{2042--2046}.
\newblock


\bibitem[\protect\citeauthoryear{Spearman}{Spearman}{1904}]%
        {Spearman04-TPA}
\bibfield{author}{\bibinfo{person}{Charles Spearman}.}
  \bibinfo{year}{1904}\natexlab{}.
\newblock \showarticletitle{{The Proof and Measurement of Association between
  Two Things}}.
\newblock \bibinfo{journal}{\emph{The American Journal of Psychology}}
  \bibinfo{volume}{15} (\bibinfo{year}{1904}), \bibinfo{pages}{72--101}.
\newblock


\bibitem[\protect\citeauthoryear{Whiting and Donthu}{Whiting and
  Donthu}{2006}]%
        {Whiting06-MVE}
\bibfield{author}{\bibinfo{person}{Anita Whiting} {and} \bibinfo{person}{Naveen
  Donthu}.} \bibinfo{year}{2006}\natexlab{}.
\newblock \showarticletitle{{Managing voice-to-voice encounters: reducing the
  agony of being put on hold}}.
\newblock \bibinfo{journal}{\emph{Journal of Service Research}}
  \bibinfo{volume}{8} (\bibinfo{year}{2006}), \bibinfo{pages}{234--244}.
\newblock


\end{thebibliography}

\end{document}